# An active efficient coding model of optokinetic nystagmus

**Chong Zhang** (czhangab@connect.ust.hk)   Department of Electrical and Computer Engineering, Hong Kong University of Science and Technology, Hong Kong

**Jochen Triesch** (triesch@fias.uni-frankfurt.de)   Theoretical Life Sciences, Frankfurt Institute for Advanced Studies, Germany

**Bertram E. Shi** (eebert@ust.hk)   Department of Electrical and Computer Engineering, Hong Kong University of Science and Technology, Hong Kong

## Abstract

Optokinetic nystagmus (OKN) is an involuntary eye movement responsible for stabilizing retinal images in the presence of relative motion between an observer and the environment. Fully understanding the development of optokinetic nystagmus requires a neurally plausible computational model that accounts for the neural development and the behavior. To date, work in this area has been limited. We propose a neurally plausible framework for the joint development of disparity and motion tuning in the visual cortex and of optokinetic and vergence eye movement behavior. To our knowledge, this framework is the first developmental model to describe the emergence of OKN in a behaving organism. Unlike past models, which were based on scalar models of overall activity in different neural areas, our framework models the development of the detailed connectivity both from the retinal input to the visual cortex, as well as from the visual cortex to the motor neurons. This framework accounts for the importance of the development of normal vergence control and binocular vision in achieving normal monocular OKN (mOKN) behaviors. Because the model includes behavior, we can simulate the same perturbations as performed in past experiments, such as artificially induced strabismus. The proposed model agrees both qualitatively and quantitatively with a number of findings from the literature on both binocular vision as well as the optokinetic reflex. Finally, our model also makes quantitative predictions about OKN behavior using the same methods used to characterize OKN in the experimental literature.

Keywords: active efficient coding, optokinetic nystagmus, monocular OKN asymmetry, binocular vision

## 1. Introduction

Perception and behavior are intricately linked. Deficits in one often lead to deficits in the other. Despite what might conceptually appear to be a potentially fragile co-dependency, in most cases, both develop robustly and in tandem. Computational models of the joint development of perception and behavior in biological systems will lead not only to a better understanding, but also possibly to corrective interventions when the processes goes awry, as well as artificial robotic systems that exhibit robust and adaptive behavior in uncertain and nonstationary environments.

Here, we describe the application of one model of joint development, the active efficient coding framework, to model the optokinetic response. The active efficient coding framework extends Barlow's efficient coding hypothesis (Barlow, 1961), to include behavior. It posits that not only do neurons in the brain develop to encode the sensory stimulus efficiently, but that behavior develops to shape the sensory input so that it can be efficiency encoded. The optokinetic reflex is an involuntary eye movement that stabilizes retinal images in the presence of relative motion between an observer and the environment. The eyes move to minimize retinal slip, the difference between the stimulus velocity projected onto the retina and the rotational velocity of the eye. In the presence of constant relative motion, the optokinetic response leads to a repetitive movement of the eyes known as the optokinetic nystagmus (OKN). OKN consists of two phases: a slow phase and a fast phase. During the slow phase, the eyes move smoothly to stabilize the retinal image. The fast phase is a saccadic eye movement that is triggered based on a complex interaction of the eye position, eye velocity and stimulus motion (Waddington & Harris, 2012, 2013), but is generally in the opposite direction. This type of nystagmus is often observed when looking sideways out of a moving vehicle, when it is called "railway nystagmus."

During binocular viewing, horizontal OKN in humans is normally symmetric from birth: it can be elicited in both the temporal-to-nasal (TN) and the nasal-to-temporal (NT) directions. However, during monocular viewing, horizontal OKN is asymmetric for infants younger than three months. The monocular OKN (mOKN) can be elicited in infants for stimuli moving in the TN direction, but not in the NT direction (Atkinson, 1979; Naegele & Held, 1982). The mOKN eventually becomes symmetric, but this is highly dependent upon the development of binocularity. For individuals with a developmental failure in binocularity, e.g. due to early strabismus (crossed eyes) or amblyopia (lazy eye), the mOKN asymmetry persists into adult life (Crone, 1977; Braddick & Atkinson, 1981b; Tychsen, 1993).

At birth, the mOKN is asymmetric because OKN is mediated by a subcortical pathway, which is monocular. This pathway passes through the nucleus of the optic tract and the dorsal terminal nucleus, which we abbreviate by NOT and DTN respectively (Hoffmann, 1981, 1986). The NOTs in the two hemispheres are directionally asymmetric in actuation. The left NOT drives the eyes to rotate leftward and the right NOT drives rightward rotation. Visual neurons in the left NOT are directly excited only by input from the right eye and vice versa for the right NOT. Thus, right eye input can only trigger leftward rotation and left eye input can only trigger rightward rotation. This motion is in the TN direction for both eyes. The transition from asymmetric to symmetric mOKN is thought to reflect the development of an indirect binocular pathway from the visual cortex to the NOT. If binocularity does not develop normally, e.g. due to strabismus or amblyopia, the mOKN will remain asymmetric even after the cortical pathway develops (Braddick, Atkinson, & Wattam-Bell, 2003).

Fully understanding the development of the optokinetic response, and in particular the transition from asymmetric mOKN to symmetric mOKN, will require a neurally plausible computational model that simultaneously accounts for many things. First, it should model the development of ascending connections from the left and right eyes to the visual cortex and descending connections from the visual cortex to the NOT-DTN. Second, the model should also account for the emergence of eye movements, since abnormalities in these (e.g. strabismus) interfere with the development of the symmetric mOKN. Eye movements are critical, since they influence the statistics of the sensory input, which in turn influence the connectivity from the eyes to the visual cortex. Given the importance of the development of binocularity in the emergence of the symmetrical mOKN, it is important to incorporate the development of vergence eye movements, which ensure that the left and right eyes receive correlated input. Like the optokinetic response, vergence eye movements undergo developmental changes early in life. Vergence eye movements first appear in infants at around 1 month (Aslin, 1977), and there is a significant improvement at 4 months of age (Mitkin & Orestova, 1988).

To date, work in this area has been limited. The only quantitative models for the effect of monocular versus binocular visual input we are aware of were presented by Hoffmann (1982) and by Kiorpes et al. (1996). To our knowledge, quantifying the effect of behavior has not been addressed at all. Hoffman described a simple model in which the activation in the left and right NOT were represented by single scalar values. This activation consisted of two components, a direct monocular input and a binocular input from cortex, and was stimulus direction dependent. Hoffman showed that the differences between the model NOT activation values were comparable to the measured gain of OKN in cat under a variety of conditions. Kiorpes et al. modeled the overall strength of connections from the two hemiretinae of the two eyes to the middle temporal (MT) area and from MT to the cortical pursuit systems. Although OKN and pursuit are distinct behaviors, they exhibit similar asymmetries due to strabismus. Based on their estimates of the weight parameters from measurements of pursuit and recordings from MT, Kiorpes et al. inferred that the source of the asymmetry is in the connections from MT to the cortical pursuit system, rather than the deficits in the visual motor processing.

Although these models give insight into the general sources of asymmetry, they have a number of shortcomings. First, they adopt very coarse representations of neural activity, where the activations of entire brain areas and connections between them are represented by a single scalar value. Second, they are not developmental, as gain parameters are fixed and do not evolve. Thus, they cannot explain what drives the developmental changes. Third, they cannot be applied to realistic visual input such as images. Finally, it is not obvious how they could be modified to include behavior.

This paper describes a quantitative model for the joint development of disparity and motion tuning in the visual cortex, the optokinetic response and vergence eye movements. This model takes as inputs image sequences sensed by the left and right eyes as the organism behaves within its environment.

These image sequences are affected by both motion in the environment, e.g. object motion, and motion by the observer, e.g. eye movements. The model accounts for the development of ascending connections from the eyes to the visual cortex, and their dependence upon the statistics of visual signals. It also models visually driven eye movements, including the subcortical control of OKN, and the development of cortical control of OKN and vergence eye movements. This development takes place as the organism behaves within the environment.

This work has several contributions. First, to our knowledge, it is the first developmental model to describe the emergence of OKN in a behaving organism. The model accounts for the development of both sensory processing and motor action. Second, it is the first model of OKN development that can be applied directly to input stimuli used experimentally to quantify OKN behavior, and in particular the asymmetry of the monocular OKN. This enables us to compare the behavior exhibited by model directly to the results in the experimental literature. Our model agrees both qualitatively and quantitatively with a number of findings on the degree of mOKN asymmetry after normal and strabismic development. Third, it is the first model of OKN development that explicitly takes into account the effect of vergence commands, and that can model normal and abnormal development of vergence commands and their effect the mOKN. Finally, it results in testable predictions about the quantitative changes in mOKN asymmetry under different developmental conditions.

## 2. Model Description

This section describes our developmental model for OKN. The first subsection gives an overview of the active efficient coding framework, which we use to model the joint development of behavior and eye movement control. The second subsection reviews current knowledge regarding the neural pathways controlling OKN, and hypotheses about their development. The final subsection describes how we model the neural pathway and its control of behavior. The overall structure of our model is consistent with previously proposed models (Hoffmann, 1983; Kiorpes et al., 1996; Tychsen, 1999; Masseck & Hoffmann, 2009). The discussion in this section is primarily qualitative. The appendix contains a formal mathematical description and a listing of parameter settings used in our simulations.

### 2.1. The active efficient coding framework

Our developmental model is based upon the active efficient coding framework (Zhao, Rothkopf, Triesch, & Shi, 2012; Vikram, Teulière, Zhang, Shi, & Triesch, 2014; Zhang, Zhao, Triesch, & Shi, 2014), an extension of Barlow's efficient coding hypothesis (Barlow, 1961). The efficient coding hypothesis posits that neural populations develop so that they can best represent the sensory data with as little redundancy or wasteful activity as possible. One consequence of this hypothesis is that the neural code is adapted to the statistics of the sensory input. The active efficient coding hypothesis extends this to include behavior: it posits that in addition, the organism behaves so that its sensory input can be efficiently encoded.

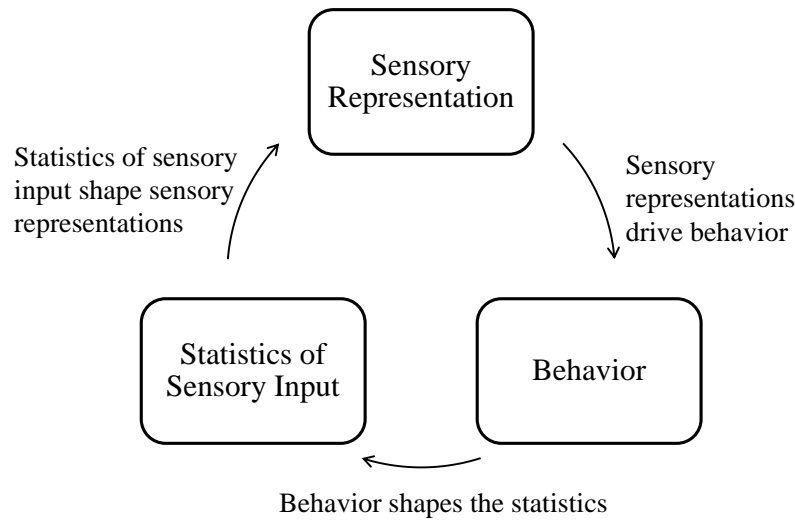

Figure 1 The active efficient coding framework.

The framework is illustrated in Figure 1. As the organism behaves in the environment, its sensory input changes. The efficient coding hypothesis predicts that the properties of the sensory neurons depend upon the statistics of this sensory input. The outputs of these sensory neurons in turn drive behavior. Under the active efficient encoding hypothesis, this behavior also develops so that the input can be efficiently encoded by the neural population. Both perception (determined by the wiring from sensory input to the sensory neurons) and behavior (determined by the wiring from sensory neurons to motor neurons) develop simultaneously as the organism behaves in the environment.

## 2.2. The neural pathway of OKN

The neural substrate of OKN is illustrated in Figure 2. The motor neurons in the left (right) NOT only drive leftward (rightward) rotation. Visual information reaches the NOT directly via retinofugal projections and indirectly via the visual cortex. The direct subcortical pathway is monocular: visual information flows from the nasal hemiretina of each eye to the contralateral NOT, as shown by the green solid line and the blue dashed line. At birth, only this subcortical pathway is functional. With monocular viewing, the optokinetic response is triggered only if the target is moving in the TN direction and only if the target is seen by the nasal hemiretina.

The cortical pathway to the NOT, which develops after birth, consists of a descending input from the ipsilateral primary visual cortex (Hoffmann, 1981, 1989). Information from the contralateral visual cortex also reaches the NOT through this descending connection via the corpus callosum. Visual input from the left (right) hemifield in both eyes is routed to the right (left) cortex. This indirect cortical pathway provides a pathway via which the NOT can receive information from the ipsilateral eye.

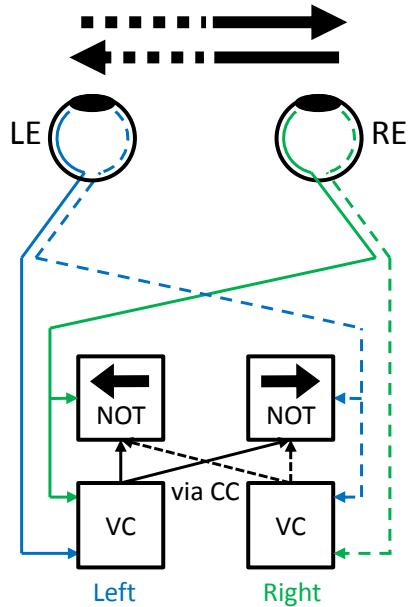

Figure 2 The neural pathway mediating OKN. Boxes with arrows indicate the nucleus of the optic tract (NOT). Boxes labelled VC indicate the visual cortex. CC: corpus callosum. LE: left eye. RE: right eye. The visual processing path is shown in blue for the left eye and green for the right eye. Solid lines indicate information from the right visual field. Dashed lines indicate information from the left visual field.

This simple routing argument is not sufficient to explain the development of the symmetric mOKN, which is highly dependent upon the development of binocularity, i.e. the combination of binocular information in the cortex. In the macaque, ocular dominance columns in the visual cortex are well formed before visual experience (Horton & Hocking, 1996), but disparity selective neurons are found a few days after birth and the spatial-frequency response properties of these neurons need several weeks to improve (Chino, Smith, Hatta, & Cheng, 1997). The development of disparity selectivity depends critically upon the eyes receiving correlated input (Hubel & Wiesel, 1965). This in turn depends upon the development of vergence control.

There have been two proposed mechanisms by which symmetry may fail to develop. One possibility is that a disruption of binocular vision may interfere with the development of the cortical connections to the NOT-DTN (Atkinson, 1979). Under this hypothesis, OKN is due to the subcortical pathway, which as pointed out above, is asymmetric. However, this hypothesis seems unlikely; since it appears that the subcortical visual pathway supporting OKN eventually disappears or becomes significantly weaker during the course of normal development (Lynch & McLaren, 1983; del Viva, Morrone, & Fiorentini, 2001). This suggests that the subcortical visual pathway serves as initial developmental scaffolding that aids in the development of the cortically driven OKN (Braddick et al., 2003). Thus, it appears more likely that the asymmetry is due to abnormalities in the descending cortical inputs to the NOT. Tychsen (1999) has suggested that cortical neurons from ocular dominance columns serving the contralateral eye preferentially drive the NOT, similar to the contralateral bias seen in the subcortical pathway. During normal binocular development, horizontal connections between ocular dominance columns emerge. These provide the pathway through which information from both the ipsilateral and contralateral eyes can reach the NOT, enabling the mOKN to become symmetric. Tychsen suggests that strabismus or amblyopia interferes with the development of these horizontal connections, leaving only input from the contralateral eye. In this case, the mOKN remains asymmetric.

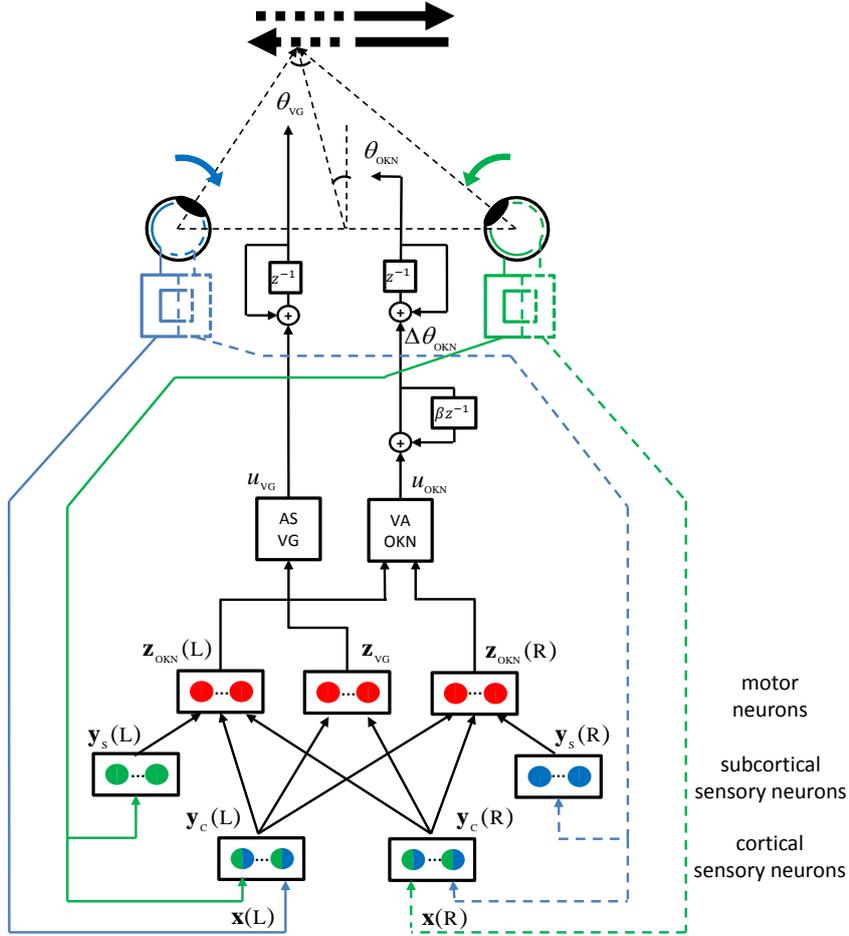

Figure 3 The developmental model the optokinetic reflex. Blue (green) lines represent the flow of information for the left (right) eye. Solid (dashed) lines represent the flow of information from the right (left) visual field. Sensory neurons are shown as blue and/or green circles. Subcortical sensory and motor neurons in the NOT are represented using solid circles. Cortical neurons are represented using half green half blue circles. Motor neurons are represented using red circles. VA: vector averaging, AS: action selection. OKN: optokinetic nystagmus. VG: vergence. $\mathbf{x}$ represents the input vector from image patches. $\mathbf{y}_s$ and $\mathbf{y}_c$ represent the subcortical and cortical sensory neurons' responses. $\mathbf{z}_{OKN}$ and $\mathbf{z}_{VG}$ represent the subcortical motor neurons' responses controlling OKN and VG. $u_{OKN}$ and $u_{VG}$ are the commands generated by the two systems. $\theta_{OKN}$ represents version angle and $\theta_{VG}$ represents vergence angle. L and R are abbreviations for the left and right hemispheres. To avoid clutter, we omit the time index.

Note that the direct subcortical and indirect cortical visual pathways we refer to here are not the same as the direct and indirect pathways governing the temporal characteristics of the OKN and OKAN, which are preserved (Cohen, Matsuo, & Raphan, 1977; Raphan, Matsuo, & Cohen, 1979; Cohen, Reisine, Yokota, & Raphan, 1992). Our model focuses upon the steady state, rather than transient, characteristics of the OKN, and thus is more consistent with the indirect or "velocity storage" mechanism, which appears to hold or store activity producing the slow phase eye velocity.

## 2.3. Developmental model of the optokinetic reflex

In this section, we give a general description of the model. The appendix describes the model in complete mathematical detail.

The detailed connectivity in our developmental model is shown in Figure 3. The model includes visually guided control of both OKN and vergence (VG). Visual information for OKN control follows both a direct monocular subcortical pathway and an indirect binocular cortical pathway. Visual information for vergence control follows only the binocular cortical pathway. The connections in the subcortical pathway are hardwired to implement OKN control. However, because this pathway is monocular, the mOKN is asymmetric. The connections in the cortical pathway, both from the retina to the cortical sensory neurons and from the cortical sensory neurons to the motor neurons controlling OKN and vergence, are initialized randomly and are learned as the agent behaves in the environment. In our simulations, we considered only horizontal eye movements, since most studies in the literature focus on horizontal, rather than vertical OKN (Knapp, Proudlock, & Gottlob, 2013). However, the model could be extended to handle vertical eye movements in a straightforward manner (Vikram et al., 2014; Zhang et al., 2014).

Our model includes inputs from both foveal and peripheral regions of the retinae (Lonini et al., 2013). The fovea is assumed to be a square region subtending 7 degrees of visual angle and centered on the optical axis. The periphery is assumed to be a square region subtending 25 degrees of visual angle and centered on the optical axis. Although there is a difference between the optical axis and the line of sight in humans, which changes over the first few months of life (Riddell, Hainline, & Abramov, 1994), we do not anticipate that this would have significant impact to the model. The changes would primarily introduce slight alterations to the geometry of image projection onto the fovea and periphery. Since the connections from retina to cortex are adaptive in our model, we expect that that our model could adapt to these differences. From a simulation point of view, the assumption that the optical axis and the line of sight coincide is convenient, since we model image formation using planar perspective projection. The difference between planar perspective projection and spherical perspective projection, which would be a better model of image formation in the eye, is smallest near the optical axis.

Both fovea and periphery are spatially sampled, but we use a coarser sampling of the periphery so that the images representing both regions are 55 by 55 pixels. Note that the periphery region includes information from the fovea region, although at a coarser scale. These images are divided into a 10 by 10 array of 10 by 10 pixel patches, which overlap by five pixels horizontally and vertically. Our model also discretizes time at a sampling rate of 20 frames per second. This sampling rate is large enough to cover the range of peak temporal frequency sensitivities in infants and adults (less than or equal to 10Hz) (Rasengane, Allen, & Manny, 1997; Dobkins, Anderson, & Lia, 1999) without temporal aliasing. Our model does not account for the effects of temporal frequency selectivity, we do not expect this to affect our results, since we do not consider OKN transients.

Processing by the left and right hemispheres of the brain is modelled separately. Patches from the left hemiretina are passed to the left hemisphere in the model and vice versa for patches from the right hemiretina. Thus, each hemisphere processes visual information from the contralateral visual hemifield.

The subcortical pathway models the sensorimotor transformations in the NOT region. They are active during the training of the cortical connections, but are disabled when the cortical pathway is tested. This is an approximation to the idea that the subcortical pathway is a scaffolding that supports the development of the cortical connections, but becomes less important later in life (Braddick et al., 2003).

Subcortical sensory neurons in the left hemisphere are assumed to respond to the retinal slip at the optic axis with Gaussian tuning curves tuned only to leftward motion in the right eye. Subcortical motor neurons in the left (right) hemisphere trigger only conjugate eye rotations in the leftward direction. The opposite is true for the neurons in the right hemisphere. There are an equal number (11) of sensory and motor neurons in each hemisphere, which are tuned to the same preferred slips and preferred rotations in units of degrees of visual angle per second. One-to-one connections between corresponding sensory and motor neurons ensure that OKN functions correctly from the start, modelling the fact that the subcortical pathway is functional at birth. However, because this pathway is monocular and neurons in each NOT only receive input from the contralateral eye, it exhibits an asymmetric mOKN as discussed earlier.

The connections from the retina to the cortical sensory neurons are learned according to a sparse coding model. The efficient coding hypothesis has been modeled mathematically by sparse coding algorithms (Olshausen & Field, 1997). Image intensities in corresponding patches from the left and right eyes and from current and previous frames are concatenated into a single input vector. Combining inputs from different eyes and different frames enables the model neurons to exhibit both disparity and motion tuning (Vikram et al., 2014). The sparse coder seeks to approximate the input vector as a sparse weighted sum of basis vectors chosen from an overcomplete dictionary. Within each scale, different patches are encoded independently, but use the same dictionary of 600 basis vectors. The left and right hemispheres are modeled separately and the fovea and periphery use separate dictionaries so that a total of four dictionaries are learned. Each basis vector is roughly analogous to the receptive field of disparity and motion tuned simple cell in the primary visual cortex. The corresponding coefficient in the weighted sum corresponds to the activation of that simple cell. We compute the responses of model complex cells by pooling the squared coefficients from the same basis vector over the different patches. This results in a population of 600 complex cells from the fovea and 600 complex cells from the periphery. During development, the basis vectors are updated to best represent the statistics of the input vectors. At each time step, they are updated to minimize the total squared reconstruction error between the input and the sparse weighted sum of basis vectors summed over all patches.

Our model assumes these cortical neurons connect to motor neurons that drive both the OKN and vergence eye movement behavior. This is consistent with findings that in adults these systems share extensive amounts of neural circuitry (Fukushima et al., 2002). These findings may not be as strong in early development. In our model, these connections are initially set to small random values and grow in magnitude during development.

The connections between the cortical sensory neurons and the motor neurons controlling OKN are learned according to a Hebbian learning rule. The bias towards connections from cortical neurons with a strong input from the contralateral eye is implemented by adding a regularizer that penalizes the size of the weights from each cortical cell to the left or right NOT according to the ocular dominance index of the cortical neurons. Connections from cortical sensory cells whose basis vectors are dominated by the ipsilateral eye are strongly penalized. Connections from cortical sensory cells whose basis vectors are balanced or dominated by the contralateral eye are favored. Motor commands to update the version angle of the two eyes are generated from the motor neuron responses by vector averaging, and are temporally smoothed by an exponential weighting function.

The connections between the cortical sensory neurons and the motor neurons controlling vergence are learned by reinforcement learning using the natural actor critic algorithm described in (Bhatnagar, Sutton, Ghavamzadeh, & Lee, 2009). The joint learning of disparity tuning and vergence control during behavior has been described in detail elsewhere (Zhao et al., 2012). Vergence commands are selected probabilistically by sampling one of the motor neurons according to a soft-max probability distribution that depends on the activation.

The vergence and version angles determined by the generated motor commands are combined to determine the left and right eye positions. If the eye positions reach a limit of motion (here set to approximately ±18 deg) a re-centering reflex is triggered to return the eyes to the center of their motion range. This re-centering reflex is a simplified version of the fast phase observed in the OKN. The actual timing, amplitude and direction of the fast phase are variable (Waddington & Harris, 2012), rather than fixed. The fast phase of OKN also appears to take on object-targeting properties like saccades (Harrison, Freeman, & Sumner, 2014, 2015). The angular range covered by the slow phase is also typically smaller than 18 degrees, which was chosen here for convenience to avoid the need to handle a large number of fast phases. Since our model is concerned primarily with the behavior in the slow phase, we do not anticipate much difference if a more accurate model of the fast phase, which would require additional modelling assumptions, is used.

# 3. Experimental Results

Note that the learning of perception and behavior takes place simultaneously in our model. Their development is mutually interdependent. Behavior changes the statistics of the input, which changes the properties of the sensory neurons. Since the sensory neurons drive the motor neurons, these changes in turn alter the behavior. Thus, it is critical to simulate not only the neural activity and plasticity within the agent, but also to place this agent within an environment.

We use the simulation environment developed for the iCub humanoid robot (Tikhanoff et al., 2008) to generate the images seen by the two eyes in response to changes in the eye positions and changes in the environment. This enables us to simulate behavior and perception in the model within a realistic, yet controlled, environment.

During training, a large planar target is placed in front of the simulated agent. The target depth is chosen randomly from a uniform distribution between 0.3 and 2 meters. The physical size of the target is fixed, but it subtends varying visual angles depending upon its depth (46 degrees at 0.3 meters, 80 degrees at 1 meter, and 141 degrees at 2 meters). A fixed background target is placed 2 meters away from the agent. Its size is chosen so that it covers 90 degrees of visual angle.

Targets move with constant horizontal speeds for periods of 1 second. During each period, a different image chosen randomly from a database of natural images (Geisler & Perry, 2011) is mapped onto the planar target. The target velocity within each period is chosen independently from a uniform distribution between -40 to 40 deg/s. The eyes' velocities are constrained to lie between -48 to 48 deg/s. Thus, possible retinal slips range from -88 to 88 deg/s.

In response to the visual input, the iCub robot makes version and vergence eye movements as described in the previous section.

During development, the subcortical pathway is active. Updates to the neural connections to the cortical sensory neurons and from the cortical sensory neurons to the motor neurons proceed in tandem with the behavior. The version angle is driven by both the subcortical and cortical sensory neurons. The vergence angle is driven only by the cortical sensory neurons.

To model the effect of strabismus, we fix the vergence angle to a constant value, independent of the vergence motor neuron output. In particular, unless otherwise noted, we introduce an esotropic strabismus, where the vergence is fixed to 20 degrees, corresponding to fixation at 0.25 meters in front of the agent when the version angle is 0 degrees.

After development, the subcortical pathway is disabled, consistent with evidence that it does not play a significant role after the cortical pathway develops (Braddick et al., 2003). The neural connections to the cortical sensory neurons and from the cortical sensory neurons to the motor neurons are fixed at their values after training. Thus, both version and vergence are driven solely by input from the same set of cortical sensory neurons.

## 3.1. Properties of the cortical sensory neurons.

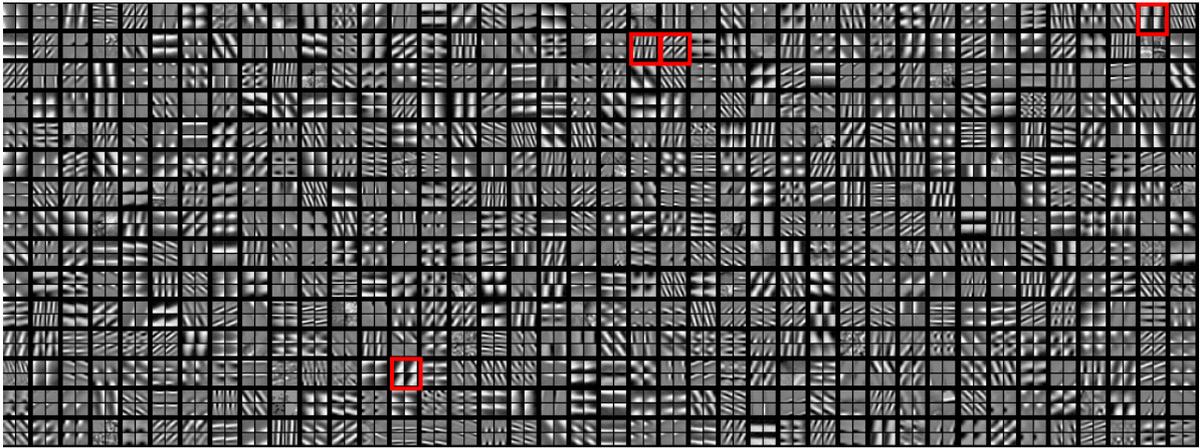

(a)

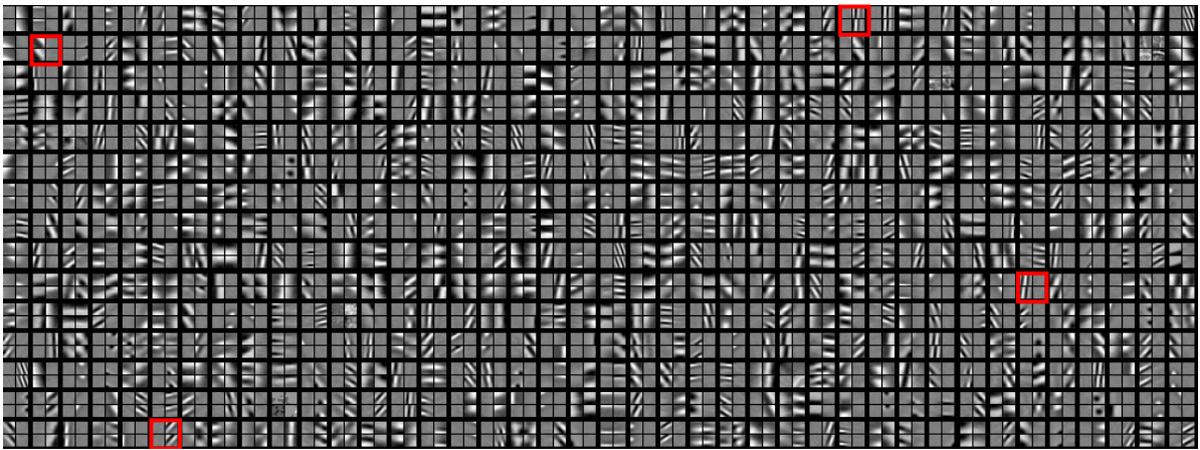

(b)

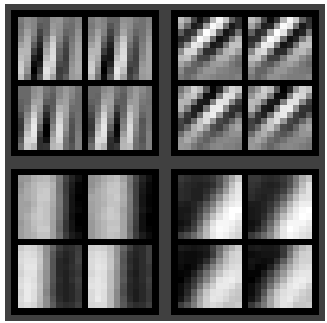          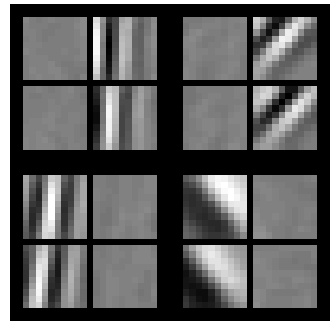

(c)                                                  (d)

Figure 4 The basis vectors of the foveal cortical sensory neurons from the left hemisphere learned under normal (a) and strabismic (b) cases. Since the input vector is a concatenation of image intensities from four patches, left/right eye at current/past frames, each basis vector is presented as four images arranged in a two by two array. Left/right columns correspond to the left/right eyes and top/bottom rows correspond to current/past frames. The 600 basis vectors are presented in a 15 by 40 array. For clarity, we show larger views of four representative basis vectors highlighted in red in (a, b) for both the normal (c) and strabismic (d) cases.

Figure 4 shows the basis vectors (analogous to receptive fields) from the cortical sensory neurons in the left hemisphere that receives input from the right eye fovea learned under both the normal and strabismic cases. The results are similar for the right hemisphere. Each basis vector is presented as a two by two array of image patches. We can study disparity tuning by comparing the patches in the left and right columns, and motion tuning by comparing the patches in the top and bottom rows. With normal development, almost all the neurons are binocular. With strabismus, most of the neurons are monocular.

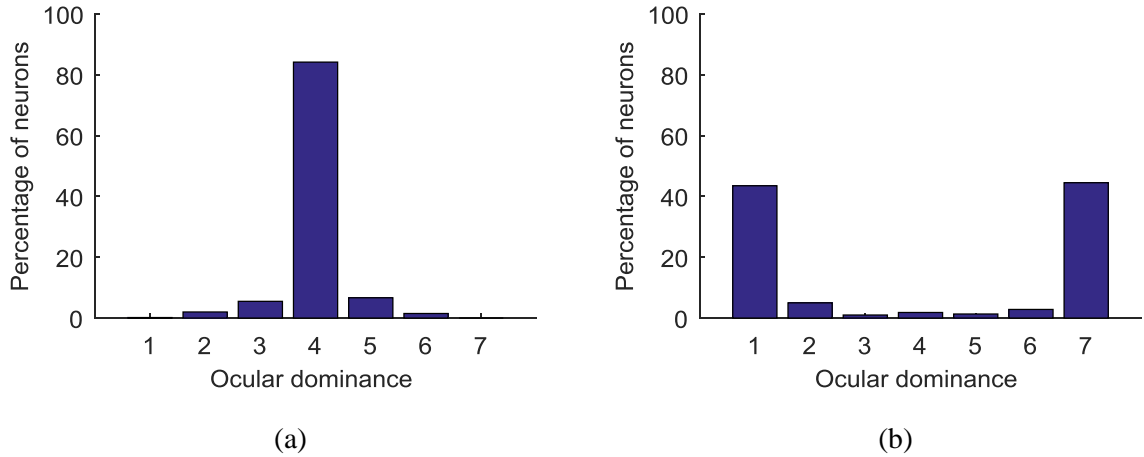

Figure 5 Ocular dominance histogram of the foveal cortical sensory neurons from left hemisphere under normal (a) and strabismic (b) cases. Neurons in bin 1 respond only to contralateral monocular input. Neurons in bin 7 respond only to ipsilateral monocular input. Neurons in bin 4 respond equally to both contralateral and ipsilateral monocular input.

We quantify the binocularity of the basis vectors by computing their ocular dominance indices (see appendix) and binning them into bins numbered from 1 for contralateral monocular to 7 for ipsilateral monocular. Bin 4 corresponds to vectors with equal weighting for both eyes. Figure 5 shows the ocular dominance (OD) histogram computed for the basis vectors shown in Figure 4. Our model produces similar results as observed by Hubel and Wiesel, e.g. Fig. 5 of reference (Hubel & Wiesel, 1965). With normal development, most neurons fall into bin 4 and there are few monocular neurons. For the strabismic case, most neurons are monocular, and only a few binocular neurons in bin 4.

For both the normal and strabismic cases, the basis vectors show similar motion tuning. To characterizing the motion tuning, we fit 2D spatial Gabor functions to the four image patches of each basis function. The Gabor functions had the same spatial frequency and Gaussian covariance matrices, but different absolute phase parameters. For the binocular neurons (with OD indices in bins 3 to 5) we fit all four image patches and computed the preferred slip from the difference between the absolute phase parameters in the top and bottom rows averaged across the two columns. For monocular neurons, we computed the fits and phase difference only for the dominant eye. Figure 6 shows the preferred slips and directions for the sensory neurons from left hemisphere (fovea) learned under (a) normal and (b) strabismic conditions from the left hemisphere. The preferred directions are almost uniformly distributed. Most preferred slips lie below 10 degrees/second.

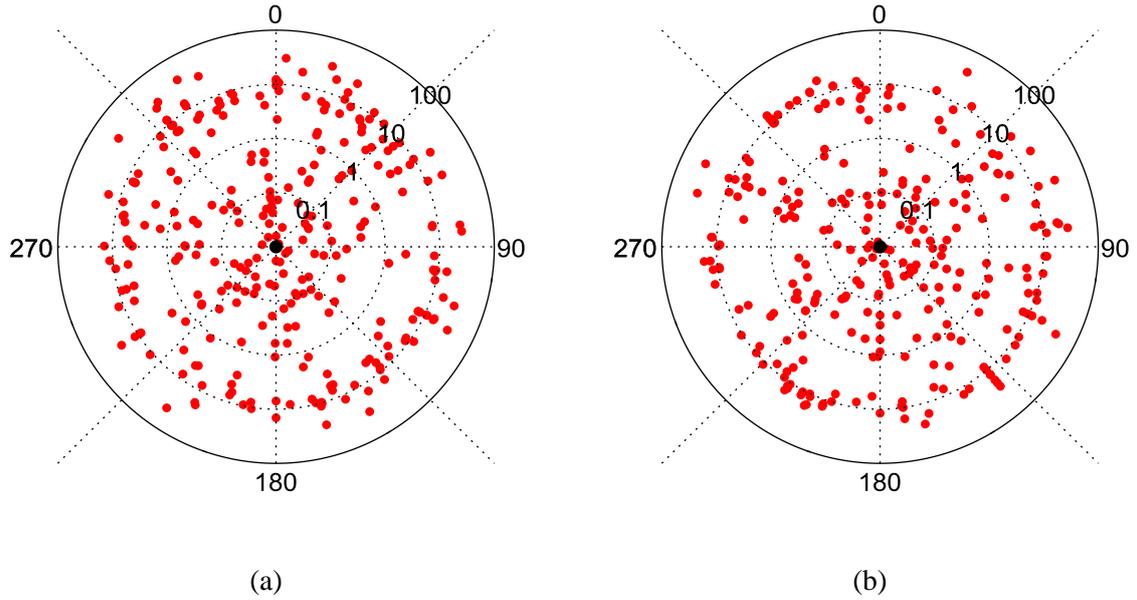

(a)                                             (b)

Figure 6 Polar plot of the distribution of preferred slips and directions for the sensory neurons from left hemisphere (fovea) learned under normal (a) and strabismic (b) condition. The angular coordinate indicates the preferred direction and the radial coordinate indicates the preferred slips in units of degrees per second. We use logarithmic scaling for the radial direction, so there is a singularity at the origin.

## 3.2. OKN Behavior

Figure 7 shows the right eye position trajectories of the agent after development under normal and strabismic conditions in response to monocular input generated by a large planar object moving with speed 30 deg/s in either the nasal-temporal (NT) or temporal-nasal (TN) direction. The object was placed 1 meter in front of the agent, and was large enough (covering 80 degrees of visual angle horizontally and vertically) so that it covered the entire visual field. Behavior for the left eye was similar.

The eye trajectories demonstrate nystagmus: there is a repeating pattern of slow and fast phases. During the slow phase, the eye tracks the object as it moves. When the eye reaches the limit of its motion (approximately ±18 deg), the fast phase returns the eye quickly to the center of its range. After normal development, the mOKN is symmetric. The curves in response to NT or TN motion are similar, except for a change in the sign of the velocity. However, with strabismus, the mOKN is asymmetric. During the slow phase, the slope for the NT stimulus is lower than the slope for the TN stimulus, indicating that the eye cannot track motion in the NT direction accurately. As a consequence, the number of fast/slow cycles (known as beats) during the same stimulus duration is lower for the NT stimulus than for the TN stimulus. The absolute rate at which fast phase occur is small compared to empirical data due to the simplifications we made in modeling the fast phase. However, since our measurement of asymmetry depends upon ratios as described below, we do not expect that a more accurate model of the fast phase to change that measures by much.

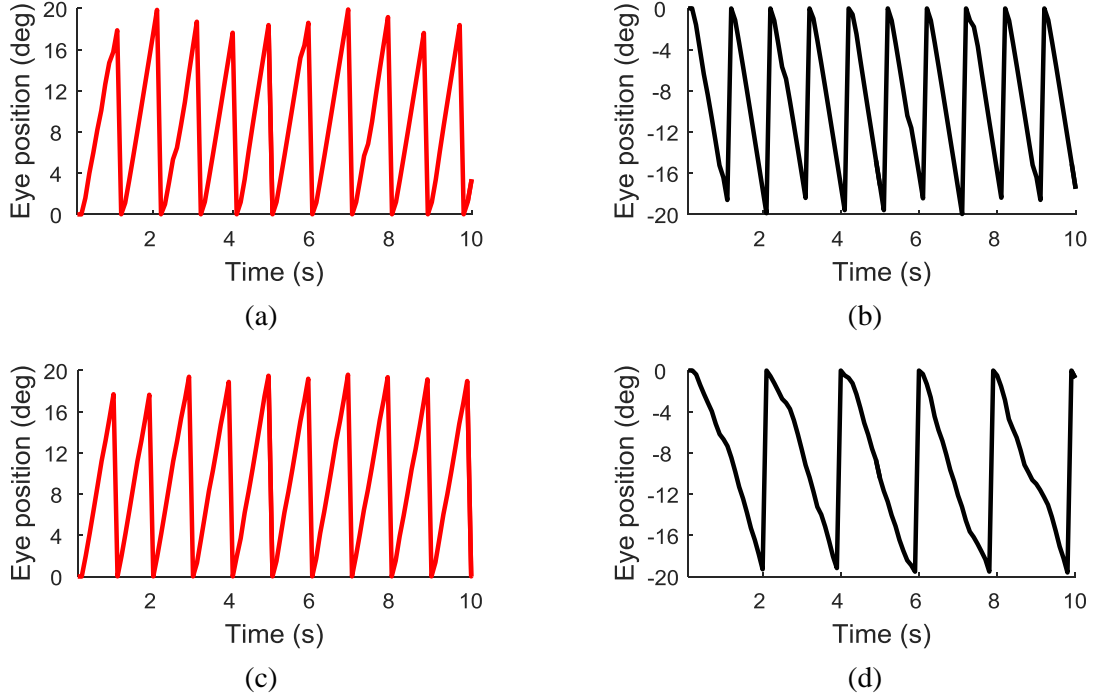

Figure 7 Eye position trajectory in response to a large planar object moving at 30 deg/s in either the temporal to nasal (a, c) or nasal to temporal (b, d) direction after development under either normal (a, b) or strabismic (c, d) conditions.

Following the experimental literature, we compute two quantitative measures for the asymmetry of mOKN behavior: the nasal bias index (NBI) (Tychsen, 2007) and the asymmetry index (ASI) (Reed et al., 1991). The NBI is based on the difference between $V_{TN}$ and $V_{NT}$, the slow phase velocities during TN and NT motion:

$$\text{NBI} = \frac{V_{TN} - V_{NT}}{V_{TN} + V_{NT}} \quad (1).$$

The NBI is zero for symmetrical mOKN, and approaches one for strongly asymmetrical mOKN. The ASI is based upon the difference between the number of beats (fast/slow cycles) for TN and NT stimuli of the same length. Denoting the number of beats for NT and TN motion by $m_{NT}$ and $m_{TN}$, we define

$$\text{ASI} = \frac{m_{TN} - m_{NT}}{m_{TN}} \quad (2).$$

The ASI is zero for symmetrical mOKN, and approaches one for strongly asymmetrical mOKN. Following (Reed et al., 1991), we will consider values of ASI greater than 0.25 to indicate asymmetric mOKN. In computing the NBI and ASI, we averaged over 100 trials in which the target moves with constant angular speed of 30 deg/s for 10 seconds.

The mOKN asymmetry after development under the strabismic condition depends upon the bias against ipsilateral connections and towards contralateral connections, which we control by a non-negative parameter *G*. Larger values of *G* create stronger biases against connections to the motor neurons from cortical neurons with significant input from the ipsilateral eye. The detailed description of the parameter is in appendix. Figure 8 shows that as the bias increases, the mOKN asymmetry increases. Tychsen tested mOKN asymmetry in primates with esotropia with visual stimuli moving at 30 deg/s, and found the NBI value to be around 0.35, matching our results with $G = 0.012$. Thus, we fixed $G = 0.012$ in the simulations reported here, including the results in Figure 7 above. The ASI computed at the end of learning is 0.05 in the normal condition and 0.50 in the strabismic condition.

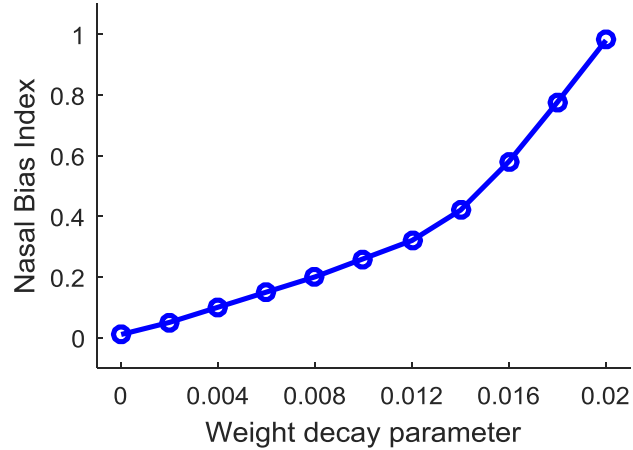

Figure 8 The mOKN asymmetry as measured by NBI increases as the parameter $G$ controlling the bias against connections to the NOT from cortical cells with strong ipsilateral eye input. The NBI was measured for the right eye for a stimulus speed equal to 30 deg/s. The curve continues to increase beyond the plotted range.

Figure 9 shows the mean slow phase velocity as a function of the stimulus speed for both normal (a) and strabismic (c) development. Eye velocity was computed using a central difference algorithm after removing the fast phase. For each velocity, the stimulus was presented for a period of 10 seconds for each trial and we repeated 100 trials. The eye velocity was computed 1 second after presenting the stimulus for each trial to remove the transient components. The mean slow phase velocity was averaged over all periods and trials. After normal development, the slow phase velocity is nearly identical for stimuli moving in both the TN and NT direction. Thus, the NBI shown in Figure 9 (b) is close to zero for all stimulus speeds. On the other hand, there is a downward shift in the slow phase velocity for the NT stimulus for the strabismic case as shown in Figure 9 (c), corresponding to the larger NBI values shown in Figure 9 (d). For slower stimuli (10 degrees/sec), the NBI was close to unity, since the slow phase velocity in the NT direction is close to zero. Similar to the results of Figure 2 of Waddington and Harris (2012), we observe a saturation in the mean slow phase velocity at higher stimulus speeds.

Figure 10 shows the population response of the motor neurons after development with strabismus. These population responses were generated by holding the eyes fixed and presenting monocular stimuli with different retinal slips to the right eye. Since the bases are normalized to have unit norm and the connections from sensory neurons to each motor neurons are also normalized (see appendix for detailed description), the motor response have arbitrary units. The scale of the response can be changed by adjusting the normalization factor. The reasons for the mOKN asymmetry is evident by comparing the magnitude of the population responses from the left NOT in response to negative (TN) slips (Figure 10 (a) in the top left) with the magnitude of the population responses from the right NOT in response to positive (NT) slips (Figure 10 (d) in the bottom right). For the right eye, the left NOT drives TN rotation and the right NOT drives NT rotation. The better stabilization for stimuli moving in the TN direction is reflected by the greater magnitude of the responses in the left NOT (Figure 10 (a)) compared with the right NOT (Figure 10 (d)). The magnitude of the responses in Figure 10 (c) and Figure 10 (b) show a similar asymmetry.

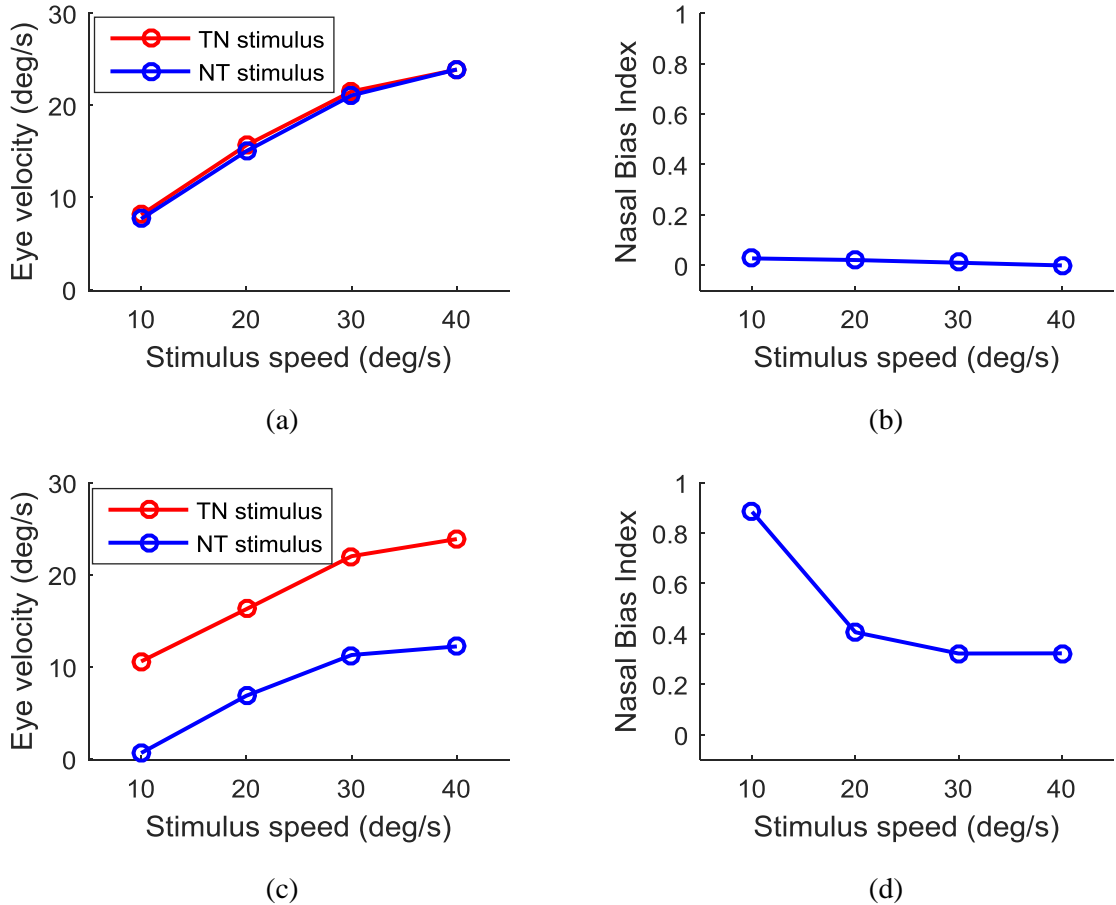

Figure 9 OKN behavior as a function of stimulus speed. The top row shows (a) the mean slow phase velocity as a function of the stimulus speed for both TN (red) and NT (blue) directed stimuli, and (b) the NBI computed for stimuli with different speeds, after normal development. The bottom row (c,d) shows the same measures after strabismic development.

Strabismus affects the statistics of the stereo disparity between the left and right eye inputs. For the development with the esotropic strabismus studied above, the distribution of stereo disparity, was biased towards large far disparities. Input in the left and right eye patches was largely uncorrelated, resulting in most basis vectors being monocular. In this case, the bias against ipsilateral connections causes the asymmetry in the cortically driven mOKN. Since the input disparity statistics determine the degree of monocularity in the basis vectors, we hypothesize that the degree of eso/exotropia disparity, can affect the degree of mOKN asymmetry by altering the proportion of inputs with near versus far disparities. Our experiments are consistent with this hypothesis. Figure 11 (a) shows that the percentage of inputs with far disparities as a function of the fixation depth corresponding to vergence angle used to simulate strabismus. The percentage decreases as the fixation depth increases, with an equal proportion of near and far disparities when the fixation depth is 1m, close to the middle of the range of target object depths. Figure 11 (b) shows that the degree of mOKN asymmetry, as measured by the NBI, reaches a minimum when the input disparities are equally distributed between near and far disparities.

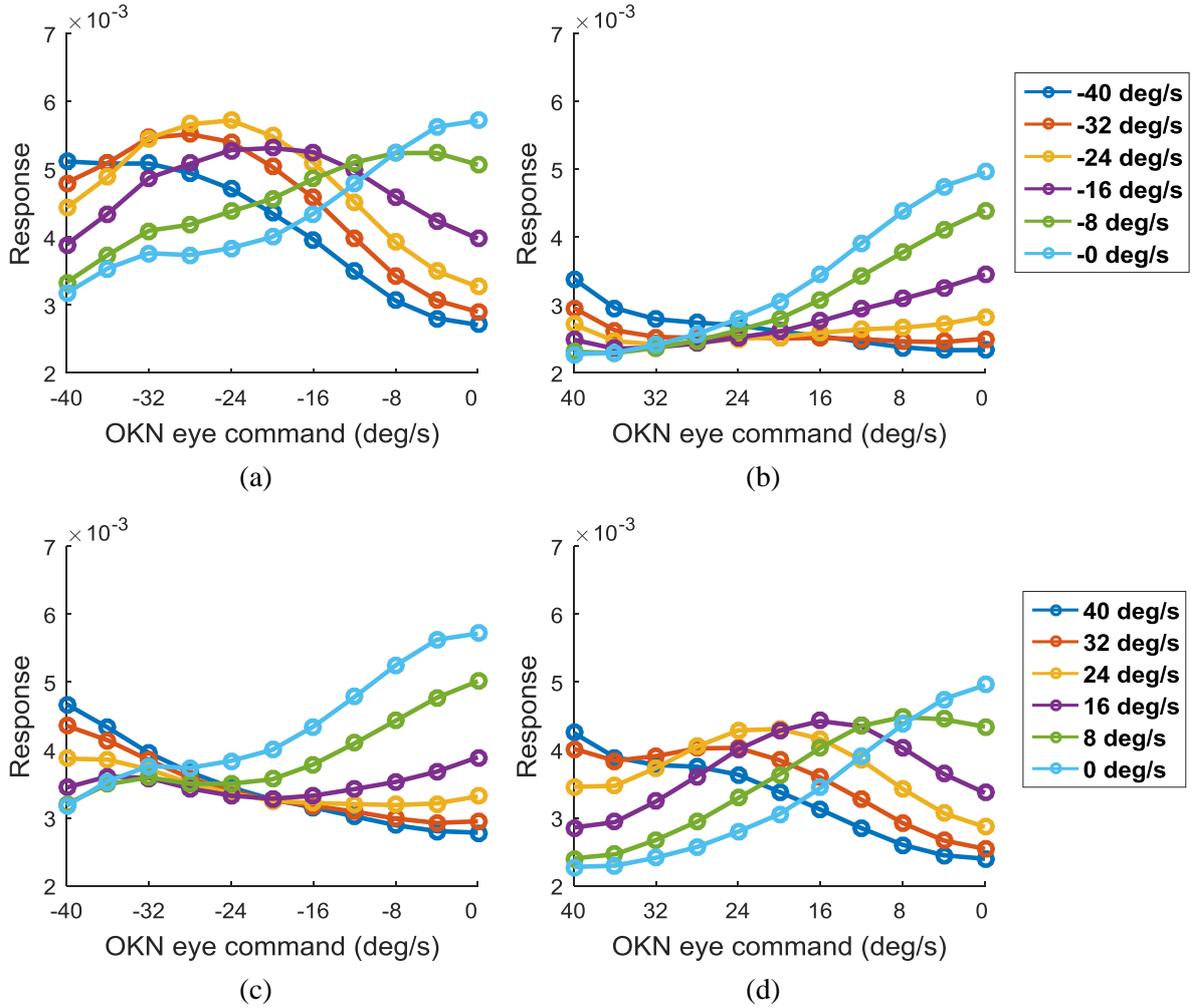

Figure 10 The population responses of the motor neurons driving version eye movements after development with strabismus in response to monocular visual stimuli presented to the right eye. Each curve shows the population response to a visual stimulus with fixed slip. Left and right columns show responses from neurons in the left and right NOT. Top and bottom rows show responses to negative and positive slips. The legend of (a) is the same as (b). The legend of (c) is the same as (d).

### 3.3. Effect of removing the contralateral bias

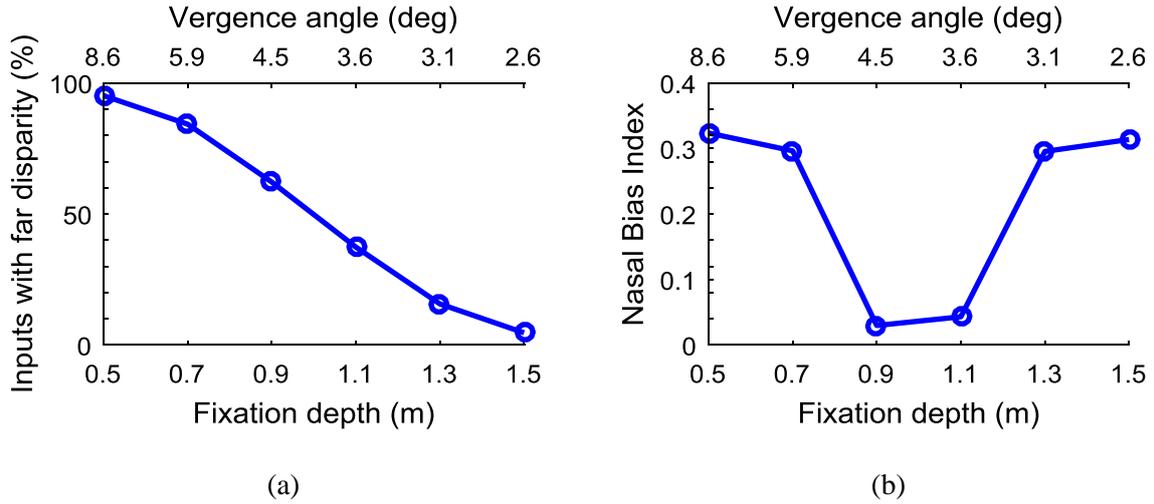

Figure 11 Behavior measure of mOKN over near-far disparity statistics. (a) Percentage of input with far disparities as a function of fixation depth (vergence angles). (b) NBI as a function of fixation depth (vergence angles). The bottom *x* axis represents the fixation depth and the top *x* axis represents the corresponding vergence angle.

In the results above, we have assumed, following (Tychsen, 1999) and others, that there is a bias towards connections to the NOT from cortical neurons dominated by contralateral retinal input. Our model enables to examine whether this assumption is necessary to account for the mOKN asymmetry observed after development with strabismus.

An alternative hypothesis for the mOKN asymmetry, which does not rely upon this bias, is that the strabismus gives rise to uncorrelated motion in the two eyes because they image independently moving surfaces. Even if all cortical neurons (whether dominated by ipsilateral/contralateral/balanced retinal input) are equally able to connect to the NOT, the connections required to stabilize motion in the NT direction will only develop under Hebbian learning if the ipsilateral eye observes NT motion at the same time as the contralateral eye is observing TN motion, since the subcortical OKN control only stabilizes TN motion in the contralateral eye. Note that TN motion in one eye corresponds to NT motion in the other eye.

We tested this hypothesis by removing the regularizer enforcing the bias against connections from cortical neurons dominated by ipsilateral input. The mOKN is symmetrical after normal development, but asymmetrical after strabismic development. The ASI computed at the end of learning is 0.04 under normal condition and 0.44 under strabismic condition. The NBI also exhibits similar trend, 0.02 for normal condition and 0.29 for strabismic condition. We simulated strabismus as in previous experiments by fixing the vergence angle to 20 degrees. The target object size was chosen so that it covered 40 degrees of visual angle at one meter. This target object size is small enough that the two eyes see different surfaces (one eye sees the target, the other the background) for a large proportion (~65%) of the time. We examine the effect of changing the target object size below.

This model predicts that the degree of mOKN asymmetry decreases with the percentage of time the two eyes observe correlated motion. This percentage will be small if the eyes image different objects. It will be large under normal development, since vergence operates correctly to ensure the two eyes fixate onto the same part of the same surface. However, the percentage can also be large with strabismus, if the planar surface is large enough that the two eyes observe the same surface. Even if the eyes image different parts of that surface, as long as the surface is undergoing fronto-parallel translation with no rotation, the motion in the two eyes will be similar.

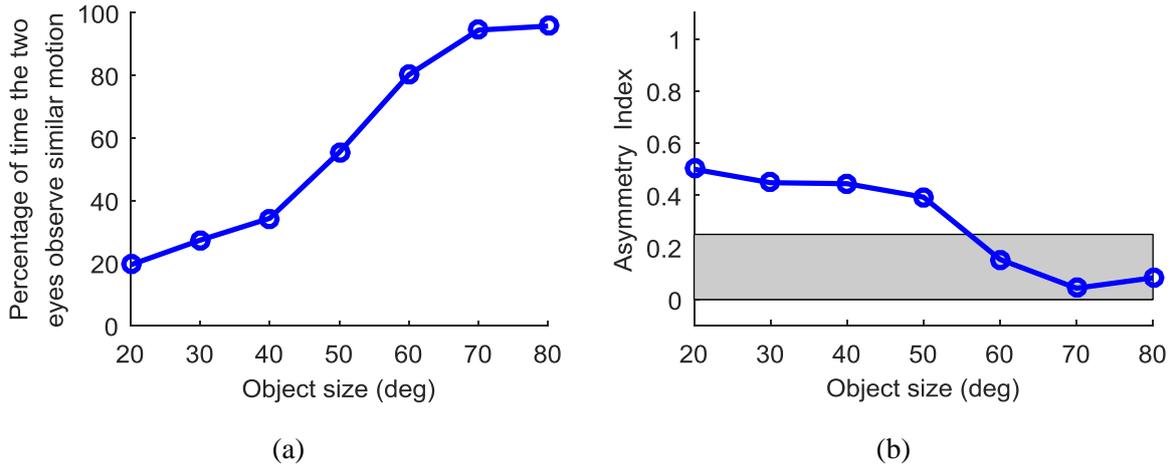

(a)                                              (b)

Figure 12 The results of strabismic development without the contralateral bias for different object sizes measured in degrees of visual angle subtended at a distance of one meter. (a) The percentage of time that the two eyes observe similar motion as the object size (defined as the visual angle subtended at a distance of one meter) changes. (b) The ASI as a function of object size. The shaded region indicates symmetric mOKN.

We tested this hypothesis by simulating strabismic development with target objects of various sizes ranging from 20 to 80 degrees of visual angle when placed at one meter distance. Figure 12 (a) plots the percentage of time the retinal slip in the two eyes was similar, as a function of object size. We defined the retinal slip to be similar if the difference was lower than one deg/s. For smaller objects, most of the time only one eye observed the moving stimulus, while the other observed the static background. As the size of the target object increased, it was seen by both eyes for a larger and larger percentage of the time. Due to the strabismus they did not necessarily observe the same part of the target, but the retinal slip in the two eyes was still similar. Figure 12 (b) shows that as the object size becomes larger, the degree of asymmetry, as measured by the ASI, decreases. If the two eyes observe the same slip for at least 80% of the time, the mOKN is symmetrical.

## 4. Discussion

We have described a neurally plausible framework for modelling the development of the cortical pathway driving the optokinetic reflex. This framework models the joint emergence of both perception and action, and accounts for the importance of the development of normal vergence control and binocular vision in achieving a symmetric mOKN. The framework is a based on the active efficient coding model (Zhao et al., 2012; Zhang et al., 2014; Teulière et al., 2015), which posits that neural coding and behavior both develop to ensure that the neural population can represent the input stimulus with high fidelity while requiring only a few active neurons. At the start of the simulation, OKN control is established via the subcortical pathway, which guides the learning of the cortical pathway. However, vergence control is random at the start, since it depends upon the connections from retina to sensory cortical neurons, and from the sensory cortical neurons to vergence motor neurons, both of which are initialized randomly. Thus, in our model, binocularity, disparity selectivity and vergence/version control develop simultaneously. Their development is mutually interdependent.

Unlike past models, which were based on scalar models of the overall activity in different neural areas, our framework models the detailed connectivity from the retinal input to the cortex, as well as from the cortex to the motor neurons driving version and vergence behavior. To our knowledge, this is the first model that can make quantitative predictions about behavior using the same stimuli used to characterize OKN in the experimental literature. In addition, it is also the first model to explicitly model development as well as the interaction between vergence and version control in the development of the optokinetic reflex.

This model matches both quantitatively and qualitatively with a wide range of experimental findings from the literature on both binocular vision and the optokinetic reflex. Because it includes behavior, we can simulate the same perturbations as performed in past experiments, such as artificially induced strabismus.

For example, the model matches with Hubel and Wiesel measurements of ocular dominance histograms of normal kittens and kittens with artificially introduced squint (Hubel & Wiesel, 1965). Hubel and Wiesel noted that 79% of the recorded cells were monocular (belonging to bins 1 and 7) for kittens with squint, compared to 20% in normal animals. For our model, around 84% of cells are monocular after development with strabismus, versus 2% for normal development. There is a much lower percentage of monocular cells observed in our model under normal development, because the statistics of input disparity are much more clustered around zero disparity than we would expect for normal behaving animals. For convenience, the model simulations have the agent continually looking at and verging onto large planar objects. However, a more realistic environment would have smaller objects with more complex depth contours, leading to a wider diversity of disparities. We expect that this would lead to more units with unbalanced ocular input after normal development.

Reed et al. (1991) compared 4 groups of subjects: early onset of strabismus (before 24 months of age), late onset of strabismus (after 24 months of age), monocularly enucleated subjects and a normal control group. Only subjects with early onset of strabismus showed statistically significant horizontal mOKN asymmetry, whereas other groups had symmetrical mOKN response. In Figure 1 Reed et al. (1991) used an asymmetry index (ASI) to measure the asymmetry of the mOKN. For normal subjects, the ASI were distributed within the range -0.3 to 0.25. For patients with early onset (<24 months) of strabismus, the asymmetry scores fall within the range 0.25 to 1. In our simulations, we observed similar ranges of the ASI after normal and strabismic development for the models both with (ASI equals 0.05 under normal and 0.50 under strabismic conditions) and without (ASI equals 0.04 under normal and 0.44 under strabismic conditions) the contralateral bias.

Tychsen (2007) used the nasal bias index (NBI) to measure mOKN asymmetry. In his experiments, he measured NBI values around 0.35 for strabismic subjects at stimulus speeds of 30 deg/s. We have chosen our parameters in the contralateral bias model to match this value ($G = 0.012$ in Figure 8). Our model predicts that even with this contralateral bias, the mOKN can be symmetric after strabismus if the proportion of near and far disparities observed during development is balanced. For the model without the contralateral bias, the NBI after strabismic development is around 0.3 when the two eyes usually observe uncorrelated motion.

A similar asymmetry after strabismic development has also been observed in the smooth pursuit system. Our results are consistent with experimental findings there as well. Kiorpes et al. (1996) found strong asymmetry favoring temporal-nasal stimuli in monkeys with artificially induced strabismus. Based on single-unit recordings from middle temporal (MT) neurons in strabismic monkeys, they found most MT cells to be monocular. We find strabismus to have a similar effect on the cortical sensory neurons that develop in our model (Figure 5 (b)), although the cortical neurons in our model are more similar to V1, rather than MT, neurons. They found the direction preferences in MT to be uniformly distributed, with no bias favoring nasalward motion. We also observe a nearly uniform distribution of preferred motion directions in our model for both normal and strabismic development (Figure 6). Thus, our model lends further support to Kiorpes et al. suggestion that asymmetries in visual tracking due to strabismus are not caused by asymmetries in the lower level cortical representation of visual motion, but rather in the mapping of this sensory representation to eye movements.

Kiorpes et al. (1996) also proposed a model enabling them to estimate the strength of connections from the two hemiretinae of the two eyes to left and right MT, and from MT to the higher parts of the cortical pursuit system (CPS). Based on pursuit measurements from strabismic monkeys, they estimated an enhancement in weights connecting MT to contralateral CPS and a reduction in weights connecting MT to ipsilateral CPS. Consistent with this, in the contralateral bias model with strabismus, we found the L2-norm of the weights from each cortical area to the contralateral NOT to be 1.3 times greater than the L2-norm of the ipsilateral weights. Note that we cannot make a quantitative comparison between the weights estimated by Kiorpes et al. and the weights in our model. In Kiorpes et al., the weights are scalars representing the strength of connections between entire areas. In our model, the weights are matricies representing the detailed connectivity between individual units in sensory and motor areas. In our model, the final motor command depends not only upon the size, but also upon the pattern, of these weights. In the simplified scalar model of Kiorpes et al., the two effects are confounded.

Waddington and Harris (2012) showed that the eye velocity during the slow phase increases as stimulus speed increases, but saturates at high stimulus speeds. For normal human subjects, when the stimulus speed is 10 deg/s, the slow phase eye velocity is around 8 deg/s. As the stimulus speed increases to 40 deg/s, the eye velocity only reaches around 20 deg/s. Our model exhibits similar quantitative behavior (Figure 9 (a)). At the stimulus speed of 10 deg/s, the eye velocity is 8 deg/s. When stimulus speed increases to 40 deg/s, the eye velocity reaches 23 deg/s. Waddington and Harris only tested healthy subjects. Our model predicts that strabismic subjects will exhibit quantitatively similar behavior for stimuli moving in the TN direction, but with lower eye velocities when the stimulus moves in the NT direction. Figure 9 (c) shows that for stimuli moving in the TN direction, the eye velocity shows a similar behavior for normal subjects, increasing from 11 deg/s to 24 deg/s when the stimulus speed increases from 10 deg/s to 40 deg/s. For stimuli moving in the NT direction, the eye velocity increases from around 0 deg/s to only 13 deg/s.

Our model also enables us to make other testable predictions about the effect of the input statistics on the degree of mOKN asymmetry. The contralateral bias model predicts that the percentage of near and far disparities seen will affect the degree of mOKN asymmetry, with less asymmetry when the percentages are balanced. The model without the contralateral bias predicts that the degree of the mOKN asymmetry will decrease as the percentage of time that the retinal slip in the two eyes is similar increases.

Moving forward, the modelling framework described here can be extended in a number of directions. In addition to the effect of strabismus studied here, we could study the effect of amblyopia. For example, Westall and Schor (1985) investigated the mOKN response in amblyopic patients to stimuli at different retinal locations: central, peripheral, nasal and temporal. We could also model this in our framework. Another interesting question is whether using reinforcement, rather than Hebbian, learning for OKN response results in any differences in performance. Our model has focused on the behavior during the slow phase of nystagmus. We used a simplified hard-wired re-centering reflex to account for the fast phase of nystagmus, but it would be interesting to study the development of this phase as well. Instead of using full field stimulus, another possible extension is to apply the framework to small target tracking task, like smooth pursuit eye movements.

## 5. Conclusion

We have proposed a neurally plausible framework for modelling the development of the cortical pathway driving the optokinetic reflex. The framework models the joint development of both perception and action. Unlike past models, which were based on scalar models of the overall activity in different neural areas, our framework models the detailed connectivity from the retinal input to the cortex, as well as from the cortex to the motor neurons driving version and vergence behavior. The proposed model agrees both qualitatively and quantitatively with a number of findings from the literature on both binocular vision as well as the optokinetic reflex. Our model also makes quantitative predictions about OKN behavior using the same methods used to characterize OKN in the experimental literature.

## 6. Acknowledgements

This work was supported in part by the Hong Kong Research Grants Council under grant number 618512 and the German BMBF under grants 01GQ1414 and 01EW1603A. JT was supported by the Quandt foundation.

## Appendix A

This appendix describes the mathematical details of the model and the simulation environment we used to perform our experiments.

### A1.  Retinal Processing

For the purposes of simulation, retinal images acquired by the left and right eye are sampled in both space and time. The temporal sampling rate is 20 frames per second.

Spatially, we model information flow from both fovea and periphery. The fovea region is assumed to be a square window covering 7 degrees of visual angle and centered on the optical axis. The peripheral region is assumed to be a square window covering 25 degrees of visual angle and centered on the optical axis. The fovea is sampled at 7.8 pixels per degree and the periphery at 2.2 pixels per degree, resulting in 55 by 55 pixel images for both cases. These images are further divided into a 10 by 10 array of 10 by 10 pixel patches. The overlap between neighboring patches is 5 pixels horizontally and vertically. The 100 patches are divided to two sets of 50, corresponding to the left and right hemiretinae. Information from the right (left) hemiretina is routed to the right (left) subcortical and cortical hemispheres. The sizes of the patches (1.3 degrees in the fovea and 4.5 degrees in the periphery) are comparable to receptive field sizes measured in V1 (Gattass, Gross, & Sandell, 1981). However, the spatial resolution is much lower than that of the cones in the retina (Yuodelis & Hendrickson, 1986). This suggests that the spatial structure of the learned receptive fields in our model will be coarser than in biology, but should still capture their essential characteristics. Simulating the connectivity at the resolution of the biological retina would have been computationally prohibitive.

Corresponding patches from the left and right eye and at sample indices $t$ and $t$-1 are concatenated into 400 dimensional input vectors, denoted by $\mathbf{x}(h,k,n,t)$, where $h \in \{\text{L,R}\}$ indexes the hemisphere, $k \in \{\text{F,P}\}$ indexes the region (Fovea or Periphery), $n \in \{1,\ldots,50\}$ indexes the patch and $t$ indexes time. The input vectors are normalized to have zero mean and unit variance.

## A2. Subcortical Sensory Processing

Since our model is primarily concerned with the development of the cortical pathway driving OKN, and the subcortical pathway eventually loses its influence, we do not model the details of the subcortical sensory processing. We assume that the population of sensory neurons in the NOT responds to the horizontal component of the retinal slip at the optical axis, which can be computed in our simulations given the object velocity, eye velocity and the imaging geometry. The neurons have Gaussian tuning curves. Their response is given by

$$y_{s,i}(h,t) = A \cdot \exp\left[-\frac{\left(\varepsilon(t) - \hat{\varepsilon}_i(h)\right)^2}{2\sigma^2}\right] \qquad (3).$$

where $h \in \{L, R\}$, $t$ indexes time, $i = \{1, ..., 11\}$ indexes the neuron, $\varepsilon(t)$ is the retinal slip at time $t$, $\hat{\varepsilon}_i(h)$ is the preferred slip, and the parameters $A$ and $\sigma$ determine the height and width of the tuning curve. Since sensory neurons in the left (right) NOT are tuned to leftward (rightward) motion, we set the preferred slips of the sensory neurons to be equally spaced from -40 to 0 deg/s for the left NOT and equally spaced from 0 to 40 deg/s for the right NOT. We set $A = 0.1$ and $\sigma = 4$ deg/s. We concatenate the responses from the sensory neurons into a single vector denoted by $\mathbf{y}_s(h,t) = \{y_{s,i}(h,t)\}_{i=1}^{11}$.

## A3. Cortical Sensory Processing

We use a sparse coding algorithm to model the outputs of the cortical neurons. Our model treats the input from each patch from each scale and each hemisphere and at each time sample identically and independently. Each input vector $\mathbf{x}(h,k,n,t)$ is approximated as the sparse weighted sum of unit norm basis vectors taken from an over-complete dictionary $\boldsymbol{\phi}_i(h,k,t)$ where $i \in \{1, ..., 600\}$. The two hemispheres ($h$) and foveal/peripheral regions ($k$) have different dictionaries, which evolve over time. The approximation is given by

$$\mathbf{x}(h,k,n,t) \approx \sum_{i=1}^{600} \alpha_i(h,k,n,t) \boldsymbol{\phi}_i(h,k,t) \qquad (4).$$

We use the matching pursuit algorithm proposed by Mallat and Zhang (1993) to choose the coefficients $\{\alpha_i(h,k,n,t)\}_{i=1}^{600}$ such that the reconstruction error,

$$e(h,k,n,t) = \left\|\mathbf{x}(h,k,n,t) - \sum_{i=1}^{600} \alpha_i(h,k,n,t) \boldsymbol{\phi}_i(h,k,t)\right\|^2 \qquad (5).$$

is small and at most 10 of them are nonzero.

The dictionaries of basis vectors $\{\boldsymbol{\phi}_i(h,k,t)\}_{i=1}^{600}$ evolve over time so that they best represent the statistics of the input patches. At each time step, we update the basis vectors using an online two step procedure similar to that used by Olshausen and Field (1997). In the first step, we find the coefficients $\alpha_i(h,k,n,t)$ using matching pursuit. In the second step, we assume the coefficients are constant, and update the basis vectors using gradient descent to minimize the total normalized squared reconstruction error over all patches

$$\sum_n e(h,k,n,t) \tag{6}.$$

After each update, the bases are re-normalized so that they are unit norm.

Each of the 600 basis vectors is roughly analogous to the receptive field of a binocular and motion tuned simple cell in the primary visual cortex. The coefficients $\alpha_i(h,k,n,t)$ are analogous to the activation of the simple cells responding to the visual information at time $t$ from patch $n$ from scale $k$ of hemiretina $h$. We model the output of complex cells by pooling the squared coefficients for each basis vector over the set of all patches:

$$y_{C,i}(h,k,t) = \frac{1}{N}\sum_{n=1}^{N} \alpha_i(k,k,n,t)^2 \tag{7}.$$

where $N = 50$. We concatenate these model outputs into a feature vector $\mathbf{y}_C(h,k,t) = \{y_{C,i}(h,k,t)\}_{i=1}^{600}$.

## A4. OKN control

OKN control is mediated by model motor neurons in the left and right NOT areas. Motor neurons in the left (right) NOT control the leftward (rightward) conjugate eye rotations in both eyes. Each NOT contains 11 motor neurons, corresponding to preferred rotations equally spaced from -40 to 0 deg/s in the left NOT and equally space from 0 to 40 deg/s in the right NOT. We choose 11 motor neurons here for simulation convenience to match the number of subcortical sensory neurons, and to match parameter settings used in previous work (Zhao, Rothkopf, Triesch, & Shi, 2012)

We assume a linear model of the motor neuron responses $\mathbf{z}_{OKN} \in \mathbb{R}^{11}$:

$$\mathbf{z}_{OKN}(h,t) = \mathbf{W}_{OKN,S} \cdot \mathbf{y}_S(h,t) + \sum_{\eta \in \{L,R\}} \sum_{k \in \{F,P\}} \mathbf{W}_{OKN,C}(h,\eta,k,t)\mathbf{y}_C(\eta,k,t) \tag{8}.$$

where $\mathbf{W}_{OKN,S} \in \mathbb{R}^{11 \times 11}$ and $\mathbf{W}_{OKN,C} \in \mathbb{R}^{11 \times 600}$ are weight matrices determining the connections from the subcortical and cortical sensory neurons to motor neurons. The first index $h$ of $\mathbf{W}_{OKN,C}$ represents the hemisphere where the motor neurons are located and the second index $\eta$ represents the hemisphere where the cortical sensory neurons located. The subcortical pathway has only ipsilateral connections, but the cortical pathway has both ipsilateral and contralateral connections.

Motor commands are generated from the motor neuron responses using vector averaging:

$$u_{OKN}(t) = \frac{\sum_{h \in \{L,R\}} \hat{\boldsymbol{\varepsilon}}(h)^T \mathbf{z}_{OKN}(h,t)}{\sum_{h \in \{L,R\}} \|\mathbf{z}_{OKN}(h,t)\|_1} \tag{9}.$$

where $\hat{\boldsymbol{\varepsilon}}(h) = \{\hat{\varepsilon}_i(h)\}_{i=1}^{11}$ is the vector of preferred rotations, which we have for convenience chosen to be the same as the vector of preferred retinal slips in the subcortical sensor neurons and the norm on the bottom is the $l_1$ norm.

The motor command is used to update the version angle of the two eyes according to

$$\theta_{OKN}(t) = \theta_{OKN}(t-1) + \Delta\theta_{OKN}(t-1) \tag{10}$$

where $\Delta\theta_{OKN}(t) = \beta \cdot \Delta\theta_{OKN}(t-1) + (1-\beta) \cdot u_{OKN}(t)$ and $\beta$ controls the time constant of an exponential smoothing filter applied to the motor commands.

The subcortical connections are hardwired and do not change over time during training. Since the preferred set of retinal slips in the sensory neurons and the preferred set of rotations in the motor neurons are identical, we choose

$$\mathbf{W}_{OKN,S} = \mu \cdot \mathbf{I} \tag{11}$$

where $\mathbf{I}$ is an identity matrix and the parameter $\mu = 10$ controls the synaptic strength. Thus, retinal slips excite corresponding eye rotations. This ensures that the subcortical pathway functions from the start to stabilize the retinal input when viewing objects moving in the temporal nasal direction. During testing, the subcortical connections are removed by setting $\mu = 0$.

The weights of the cortical connections evolve over time according to a Hebbian learning rule.

$$\mathbf{W}_{OKN,C}(h,\eta,k,t+1) = \mathbf{W}_{OKN,C}(h,\eta,k,t)\boldsymbol{\Gamma}(h,\eta,k,t) + \kappa \cdot \mathbf{z}_{OKN}(h,t)\mathbf{y}_C(\eta,k,t)^T \tag{12}$$

where $\boldsymbol{\Gamma}(h,\eta,k,t) = \text{diag}(\gamma_1(h,\eta,k,t),\ldots,\gamma_{600}(h,\eta,k,t))$ is a diagonal matrix of weight decay parameters and $\kappa$ is a positive learning rate parameter. After each update, each row of the weight vector is normalized so that the sum of the weights entering each motor neuron sums to one. Weights are initialized to small random values drawn from independent uniform distributions before normalization.

The weight decay parameters bias the motor neurons to make stronger connections with cortical neurons that contain significant input from the contralateral eye. The smaller the weight decay parameter $\gamma_i(h,\eta,k,t)$ the more strongly connections from the model complex cell $i$ in cortical hemisphere $\eta$ to the motor neurons in the NOT in hemisphere $h$ are penalized. The value of the weight decay parameter depends upon $OD_i(\eta,k,t)$, the ocular dominance index of the complex cell $i$ from hemisphere $\eta$ and region $k$, according to the equation

$$\gamma_i(h,\eta,k,t) = \begin{cases} 1 - G + \dfrac{G}{1+\exp(a(OD_i(\eta,k,t)+b))} & \text{if } h = R \\ 1 - G + \dfrac{G}{1+\exp(a(-OD_i(\eta,k,t)+b))} & \text{if } h = L \end{cases} \tag{13}$$

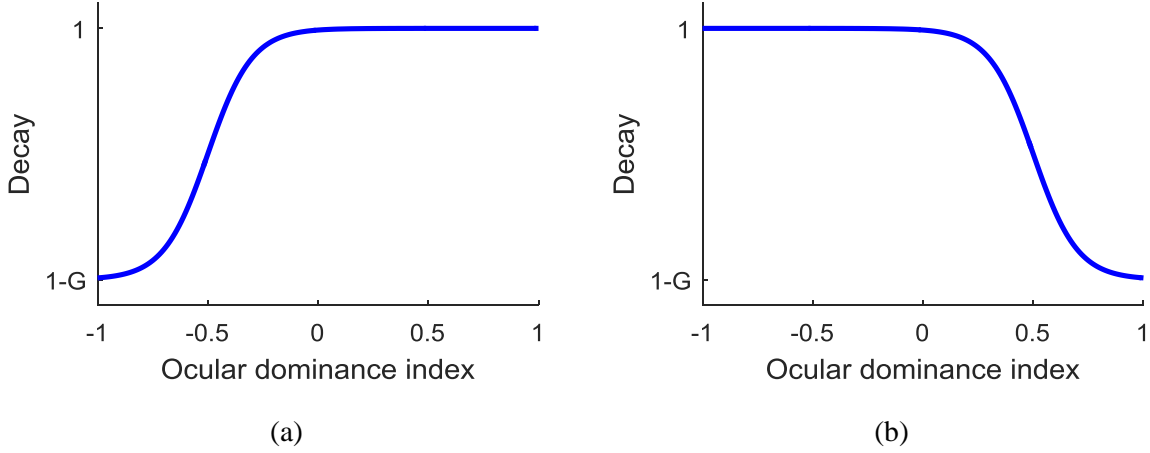

(a)                  (b)

Figure 13 The weight decay parameter as a function of ocular dominance index for the weights connecting to the right (a) and left (b) NOT.

which is plotted in Figure 13. The parameter $G$ controls the extent to which the connections are penalized. The parameter $a$ controls the slope of the transition at the threshold $+b$ ($-b$) for the left (right) NOT. We chose $a=10$ and $b=0.5$.

The ocular dominance index is computed according to the method described in Hoyer and Hyvärinen (2000). If we let $\phi_{i,L}(h,k,t)$ and $\phi_{i,R}(h,k,t)$ denote the parts of the basis vector $\phi_i(h,k,t)$ corresponding to the left and right eye inputs, then

$$\text{OD}_i(h,k,t) = \frac{\|\phi_{i,L}(h,k,t)\| - \|\phi_{i,R}(h,k,t)\|}{\|\phi_{i,L}(h,k,t)\| + \|\phi_{i,R}(h,k,t)\|} \tag{14}.$$

The OD is equal to 1 (-1) if the complex cell receives only left (right) eye input and equal to zero if it receives balanced input. The OD changes over time as the basis vectors change due to the training of the sparse coder. When computing histograms of the ocular dominance, we typically use 7 bins with boundaries [-0.85, -0.5, -0.15, 0.15, 0.5, 0.85], which is same as (Shouval, Intrator, Law, & Cooper, 1996). This facilitates comparison with prior experimental results, e.g. (Hubel & Wiesel, 1965)

## A5. Vergence Control

Vergence control is mediated by 11 (again chosen for convenience and consistency with prior work) motor neurons encoding preferred changes in the vergence angle, which are equally spaced from -1 to 1 deg. We assume a linear model for the motor neuron responses $\mathbf{z}_{VG} \in \mathbb{R}^{11}$

$$\mathbf{z}_{VG}(t) = \sum_{h \in \{L,R\}} \sum_{k \in \{F,P\}} \mathbf{W}_{VG}(h,k,t) \mathbf{y}_C(h,k,t) \tag{15}.$$

where $\mathbf{W}_{VG} \in \mathbb{R}^{11 \times 600}$ are weight matrices determining the connection from the cortical neurons to the vergence motor neurons.

The motor commands for vergence $u_{VG}(t)$ are obtained by choosing the preferred change in the vergence angle corresponding to one of the motor neurons, which is sampled from the probability distribution obtained by applying a soft-max function to the vector of motor neuron responses, $\mathbf{z}_{\text{VG}}(t) = \{z_{\text{VG},i}(t)\}_{i=1}^{11}$:

$$\pi_i(t) = \frac{\exp(z_{\text{VG},i}(t)/T)}{\sum_{j=1}^{11} \exp(z_{\text{VG},j}(t)/T)} \quad (16).$$

where $T$ is a positive temperature parameter controlling the greediness of the soft-max function. Given the vergence command, the vergence angle is updated according to

$$\theta_{\text{VG}}(t) = \theta_{\text{VG}}(t-1) + u_{\text{VG}}(t-1) \quad (17).$$

The weights are updated using reinforcement learning according to the natural actor-critic (NAC) reinforcement learning algorithm described in (Bhatnagar et al., 2009) with the addition of weight decay. The reward is the negative discounted sum of the total reconstruction error $-\sum_{h,k,n} e(h,k,n,t)$, where $e(h,k,n,t)$ is given in (5).

# 7. References

Aslin, R. N. (1977). Development of binocular fixation in human infants. *Journal of Experimental Child Psychology, 23*(1), 133-150.

Atkinson, J. (1979). Development of optokinetic nystagmus in the human infant and monkey infant: an analogue to development in kittens. In R. D. Freeman (Ed.), *Developmental neurobiology of vision. NATO advanced study institute series* (pp. 277-287). New York: Plenum Press.

Barlow, H. B. (1961). Possible principles underlying the transformation of sensory messages. In W. A. Rosenblith (Ed.), *Sensory Communication* (pp. 217-234). Cambridge, MA: MIT Press.

Bhatnagar, S., Sutton, R. S., Ghavamzadeh, M., & Lee, M. (2009). Natural actor–critic algorithms. *Automatica, 45*(11), 2471-2482.

Braddick, O., & Atkinson, J. (1981b). Development of optokinetic nystagmus in infants: an indicator of cortical binocularity? In D. F. Fisher, R. A. Monty & J. W. Senders (Eds.), *Eye movements: Cognition and visual perception* (pp. 53-64). Hillsdale, NJ: Lawrence Erlbaum Associates.

Braddick, O., Atkinson, J., & Wattam-Bell, J. (2003). Normal and anomalous development of visual motion processing: motion coherence and 'dorsal-stream vulnerability'. *Neuropsychologia, 41*(13), 1769-1784.

Chino, Y. M., Smith, E. L., Hatta, S., & Cheng, H. (1997). Postnatal development of binocular disparity sensitivity in neurons of the primate visual cortex. *The Journal of Neuroscience, 17*(1), 296-307.

Cohen, B., Matsuo, V., & Raphan, T. (1977). Quantitative analysis of the velocity characteristics of optokinetic nystagmus and optokinetic after‐nystagmus. *The Journal of Physiology, 270*(2), 321-344.

Cohen, B., Reisine, H., Yokota, J. I., & Raphan, T. (1992). The Nucleus of the Optic Tracta: Its Function in Gaze Stabilization and Control of Visual‐Vestibular Interaction. *Annals of the New York Academy of Sciences, 656*(1), 277-296.

Crone, R. (1977). Amblyopia: the pathology of motor disorders in amblyopic eyes. *Documenta Ophthalmologica Proceedings Series, 45*, 9-18.

del Viva, Morrone, M. M., & Fiorentini, A. (2001). VEP selective responses to flow motion in adults and infants. *Perception, 30*, 36.

Dobkins, K. R., Anderson, C. M., & Lia, B. (1999). Infant temporal contrast sensitivity functions (tCSFs) mature earlier for luminance than for chromatic stimuli: evidence for precocious magnocellular development? *Vision Research, 39*(19), 3223-3239.

Fukushima, K., Yamanobe, T., Shinmei, Y., Fukushima, J., Kurkin, S., & Peterson, B. W. (2002). Coding of smooth eye movements in three-dimensional space by frontal cortex. *Nature, 419*(6903), 157-162.

Gattass, R., Gross, C., & Sandell, J. (1981). Visual topography of V2 in the macaque. *Journal of Comparative Neurology, 201*(4), 519-539.

Geisler, W. S., & Perry, J. S. (2011). Statistics for optimal point prediction in natural images. *Journal of Vision, 11*(12), 14.

Harrison, J. J., Freeman, T. C., & Sumner, P. (2014). Saccade-like behavior in the fast-phases of optokinetic nystagmus: An illustration of the emergence of volitional actions from automatic reflexes. *Journal of Experimental Psychology: General, 143*(5), 1923.

Harrison, J. J., Freeman, T. C., & Sumner, P. (2015). Saccadic compensation for reflexive optokinetic nystagmus just as good as compensation for volitional pursuit. *Journal of Vision, 15*(1), 24-24.

Hoffmann, K.-P. (1981). Neuronal responses related to optokinetic nystagmus in the cat's nucleus of the optic tract. In A. Fuchs & W. Becker (Eds.), *Progress in oculomotor research* (pp. 443-454). New York: Elsevier.

Hoffmann, K.-P. (1982). Cortical versus subcortical contributions to the optokinetic reflex in the cat. In G. Lennerstrand (Ed.), *Functional basis of ocular motility disorders* (pp. 303-310). Oxford: Pergamon Press.

Hoffmann, K.-P. (1983). Control of the optokinetic reflex by the nucleus of the optic tract in the cat. In A. Hein & M. Jeannerod (Eds.), *Spatially oriented behavior* (1th ed., pp. 135-153). New York: Springer.

Hoffmann, K.-P. (1986). Visual inputs relevant for the optokinetic nystagmus in mammals. *Progress in Brain Research, 64*, 75-84.

Hoffmann, K.-P. (1989). Control of the optokinetic reflex by the nucleus of the optic tract in primates. *Progress in Brain Research, 80*, 173-182.

Horton, J. C., & Hocking, D. R. (1996). An adult-like pattern of ocular dominance columns in striate cortex of newborn monkeys prior to visual experience. *The Journal of Neuroscience, 16*(5), 1791-1807.

Hoyer, P. O., & Hyvärinen, A. (2000). Independent component analysis applied to feature extraction from colour and stereo images. *Network: Computation in Neural Systems, 11*(3), 191-210.

Hubel, D. H., & Wiesel, T. N. (1965). Binocular interaction in striate cortex of kittens reared with artificial squint. *Journal of Neurophysiology, 28*(6), 1041-1059.

Kiorpes, L., Walton, P. J., O'Keefe, L. P., Movshon, J. A., & Lisberger, S. G. (1996). Effects of early-onset artificial strabismus on pursuit eye movements and on neuronal responses in area MT of macaque monkeys. *The Journal of Neuroscience, 16*(20), 6537-6553.

Knapp, C. M., Proudlock, F. A., & Gottlob, I. (2013). OKN asymmetry in human subjects: a literature review. *Strabismus, 21*(1), 37-49.

Lonini, L., Forestier, S., Teulière, C., Zhao, Y., Shi, B. E., & Triesch, J. (2013). Robust active binocular vision through intrinsically motivated learning. *Frontiers in Neurorobotics, 7*(20), 1-10.

Lynch, J. C., & McLaren, J. W. (1983). Optokinetic nystagmus deficits following parieto-occipital cortex lesions in monkeys. *Experimental Brain Research, 49*(1), 125-130.

Mallat, S. G., & Zhang, Z. (1993). Matching pursuits with time-frequency dictionaries. *IEEE Transactions on Signal Processing, 41*(12), 3397-3415.

Masseck, O. A., & Hoffmann, K.-P. (2009). Comparative neurobiology of the optokinetic reflex. *Annals of the New York Academy of Sciences, 1164*(1), 430-439.

Mitkin, A., & Orestova, E. (1988). Development of binocular vision in early ontogenesis. *Psychologische Beitrage*.

Naegele, J. R., & Held, R. (1982). The postnatal development of monocular optokinetic nystagmus in infants. *Vision Research, 22*(3), 341-346.

Olshausen, B. A., & Field, D. J. (1997). Sparse coding with an overcomplete basis set: A strategy employed by V1? *Vision Research, 37*(23), 3311-3325.

Raphan, T., Matsuo, V., & Cohen, B. (1979). Velocity storage in the vestibulo-ocular reflex arc (VOR). *Experimental Brain Research, 35*(2), 229-248.

Rasengane, T. A., Allen, D., & Manny, R. E. (1997). Development of temporal contrast sensitivity in human infants. *Vision Research, 37*(13), 1747-1754.

Reed, M. J., Steinbach, M. J., Anstis, S. M., Gallie, B., Smith, D., & Kraft, S. (1991). The development of optokinetic nystagmus in strabismic and monocularly enucleated subjects. *Behavioural Brain Research, 46*(1), 31-42.

Riddell, P. M., Hainline, L., & Abramov, I. (1994). Calibration of the Hirschberg test in human infants. *Investigative Ophthalmology & Visual Science, 35*(2), 538-543.

Shouval, H., Intrator, N., Law, C. C., & Cooper, L. N. (1996). Effect of binocular cortical misalignment on ocular dominance and orientation selectivity. *Neural Computation, 8*(5), 1021-1040.

Teulière, C., Forestier, S., Lonini, L., Zhang, C., Zhao, Y., Shi, B. E., & Triesch, J. (2015). Self-calibrating smooth pursuit through active efficient coding. *Robotics and Autonomous Systems, 71*, 3-12.

Tikhanoff, V., Cangelosi, A., Fitzpatrick, P., Metta, G., Natale, L., & Nori, F. (2008). *An open-source simulator for cognitive robotics research: the prototype of the iCub humanoid robot simulator.* Paper presented at the Proceedings of IEEE Workshop on Performance Metrics for Intelligent Systems Workshop.

Tychsen, L. (1993). Motion sensitivity and the origins of infantile strabismus. In K. Simons (Ed.), *Early visual development: basic and clinical research* (pp. 364-390). New York: Oxford University Press.

Tychsen, L. (1999). Infantile esotropia: current neurophysiologic concepts. In A. L. Rosenbaum & A. P. Santiago (Eds.), *Clinical strabismus management* (pp. 117-138). Philadelphia Saunders.

Tychsen, L. (2007). Causing and curing infantile esotropia in primates: the role of decorrelated binocular input. *Transactions of the American Ophthalmological Society, 105*, 564.

Vikram, T. N., Teulière, C., Zhang, C., Shi, B. E., & Triesch, J. (2014). *Autonomous learning of smooth pursuit and vergence through active efficient coding.* Paper presented at the IEEE International Conference on Development and Learning and Epigenetic Robotics.

Waddington, J., & Harris, C. M. (2012). Human optokinetic nystagmus: a stochastic analysis. *Journal of Vision, 12*(12), 5.

Waddington, J., & Harris, C. M. (2013). The distribution of quick phase interval durations in human optokinetic nystagmus. *Experimental Brain Research, 224*(2), 179-187.

Westall, C. A., & Schor, C. M. (1985). Asymmetries of optokinetic nystagmus in amblyopia: the effect of selected retinal stimulation. *Vision Research, 25*(10), 1431-1438.

Yuodelis, C., & Hendrickson, A. (1986). A qualitative and quantitative analysis of the human fovea during development. *Vision Research, 26*(6), 847-855.

Zhang, C., Zhao, Y., Triesch, J., & Shi, B. E. (2014). *Intrinsically motivated learning of visual motion perception and smooth pursuit.* Paper presented at the Proc. IEEE International Conference on Robotics and Automation.

Zhao, Y., Rothkopf, C. A., Triesch, J., & Shi, B. E. (2012). *A unified model of the joint development of disparity selectivity and vergence control.* Paper presented at the IEEE International Conference on Development and Learning and Epigenetic Robotics.